\documentclass[prb,english,reprint,superscriptaddress]{revtex4-2}
\usepackage[T1]{fontenc}
\usepackage[latin9]{inputenc}
\usepackage{bm}
\usepackage{hyperref}
\usepackage{amsmath}
\usepackage{amssymb}
\usepackage{bbold}
\usepackage{babel}
\usepackage{ulem}
\usepackage{graphicx}
\newcommand{\dd}[1]{\mathrm{d}#1\,}

\newcommand{\avg}[1]{\langle{#1}\rangle}
\renewcommand{\Re}{\mathop{\mathrm{Re}}}

\DeclareMathOperator{\Tr}{Tr}
\DeclareMathOperator{\tr}{tr}

\newcommand{\Res}{\mathop{\mathrm{Res}}}

\renewcommand{\vec}[1]{\bm{#1}}

\usepackage{xcolor}

\definecolor{PV-color}{rgb}{0.97,0.57,0.11}

\definecolor{SB-color}{rgb}{0.37,0.57,0.11}

\definecolor{IT-color}{rgb}{0.57,0.11,0.97}

\newcommand{\longmath}[2][0.7\columnwidth]{\begin{minipage}[t]{#1}\raggedright\linespread{1.2}\selectfont\allowdisplaybreaks\hangindent=2em\hangafter=1\begin{math}#2\end{math}\end{minipage}}

\begin{document}

\title{Nonlinear $\sigma$-model for disordered systems with intrinsic spin-orbit coupling}

\author{P. Virtanen}
\affiliation{Department of Physics and Nanoscience Center, University of Jyv\"askyl\"a, P.O. Box 35 (YFL), FI-40014 University of Jyv\"askyl\"a, Finland}

\author{F. S. Bergeret}
\affiliation{Centro de F\'isica de Materiales (CFM-MPC) Centro Mixto CSIC-UPV/EHU, E-20018 Donostia-San Sebasti\'an,  Spain}
\affiliation{Donostia International Physics Center (DIPC), 20018 Donostia--San Sebasti\'an, Spain}
\affiliation{Institut f\"ur Theoretische Physik und Astrophysik, Universit\"at W\"urzburg, 97074 W\"urzburg, Germany}

\author{I. V. Tokatly}
\affiliation{Nano-Bio Spectroscopy Group, Departamento de Pol\'imeros y Materiales Avanzados: F\'isica, Qu\'imica y Tecnolog\'ia, Universidad del Pa\'is Vasco (UPV/EHU), 20018 Donostia-San Sebasti\'{a}n, Spain} 
\affiliation{IKERBASQUE, Basque Foundation for Science, 48011 Bilbao, Spain}
\affiliation{Donostia International Physics Center (DIPC), 20018 Donostia--San Sebasti\'an, Spain}
\affiliation{ITMO University, Department of Physics and Engineering, Saint-Petersburg 197101, Russia}

\begin{abstract}
  We derive the nonlinear $\sigma$-model to describe diffusive transport in normal metals and superconductors with intrinsic spin-orbit coupling (SOC). 
  The SOC is described via an SU(2)
  gauge field, and we expand the model to the fourth order in
  gradients to find the leading non-Abelian field-strength
  contribution. This contribution generates the spin-charge coupling that is responsible for the spin-Hall effect.
  We discuss how its symmetry differs from the leading quasiclassical higher-order gradient terms. We also derive the corresponding Usadel equation describing the diffusive spin-charge dynamics in superconducting systems. As an example,   we apply the obtained equations to describe the anomalous supercurrent in dirty Rashba superconductors at arbitrary temperatures. 
\end{abstract}

\maketitle

\section{Introduction}
\label{sec:introduction}

The intrinsic spin-Hall effect is a magnetoelectric coupling between
spin and charge degrees of freedom, which arises from a geometric property of the electron bands caused by the spin-orbit coupling (SOC). \cite{sinova2015,nagaosa2010,xiao2010-bpe}
The spin-charge interconversion is the basis of the spin-Hall magnetoresistance \cite{nakayama2013spin}, the Edelstein  \cite{song2017observation,ghiasi2019charge} and spin-galvanic effects \cite{sanchez2013spin}, observed in a wide variety of systems  \cite{meyer2014temperature,gomez2020strong,sanz2020quantification,velez2019spin}.

The counterpart of these effects in superconducting systems with SOC  has also been widely studied \cite{edelstein1995,edelstein2005magnetoelectric,yip2014noncentrosymmetric, konschelle2015theory}. Supercurrents, i.e., currents without dissipation, can induce a spin density. Reciprocally, a Zeeman or exchange field can cause supercurrents in a superconducting system with strong SOC. An example of the latter is the realization of anomalous Josephson junctions \cite{buzdin2008direct,bergeret2015theory}, where the interplay between the SOC and a spin-splitting field leads to the appearance of spontaneous supercurrents in superconducting loops \cite{strambini2020josephson,assouline2019spin}.  
The charge-spin coupling in superconducting systems with SOC is also at the basis of the superconducting diode  effect \cite{wakatsuki2018nonreciprocal,daido2022intrinsic,ilic2021effect,he2021phenomenological,davydova2022universal}, 
observed experimentally \cite{ando2020observation,baumgartner2022supercurrent}. With potential applications in emerging technologies, all these effects are observed in hybrid systems, which combine different materials with disorder. From a theoretical point of view, the formulation of a kinetic theory of electronic transport in the presence of SOC is therefore of extreme importance. 

If the system under consideration is described by an effective Hamiltonian with a  linear in momentum SOC, the latter can be treated by introducing an SU(2) gauge potential.   \cite{mineev1992,mathur1992,frohlich1993,jin2006,bernevig2006exact,tokatly2008,gorini2010}. This viewpoint  turns out to be fruitful, as the intrinsic
spin-Hall contribution in electron transport theory can be related to
the corresponding SU(2) field strength. \cite{jin2006,tokatly2008,gorini2010,shen2014,konschelle2015theory}. 
In superconductors, the magnetoelectric
contribution has been considered in various limits. \cite{edelstein1995,yip2002two,mineev2011magnetoelectric,bergeret2015theory,
konschelle2015theory,smidman2017superconductivity, edelstein2021} 
One of the questions is the effect of
impurity scattering, and the formulation of the transport theory in
terms of kinetic equations for the quasiclassical Green's functions (GFs), in  superconductors with SOC in the diffusive limit. 
This type of formulation is the most suitable for the study of realistic mesoscopic  systems, such as anomalous  Josephson junctions, superconductor-ferromagnet  or superconductor-semiconductor hybrid systems \cite{strambini2020josephson,assouline2019spin,baumgartner2022supercurrent}. 

Intrinsic magnetoelectric effects in diffusive hybrid systems have been considered  mostly in  the 
linearized case, when superconducting correlations are weak either due to large temperature or weak proximity effect. 
\cite{buzdin2008direct,bergeret2013,bergeret2015theory,konschelle2015theory}. In such a case,   the   SU(2)-covariant quasiclassical equation, the Usadel equation,  has a similar form as the diffusion equation in a normal system \cite{gorini2010} and the anomalous (superconducting) GFs can be treated perturbatively. 
Going beyond the linearized case is not a trivial task and   
attempts to obtain the Usadel equation   beyond that limit  using 
standard quasiclassical kinetic equation approaches   run into technical consistency issues \cite{tokatly2017}. 
On the  one hand the spin-Hall effect
appears in a sub-quasiclassical order in the expansions. 
On the other hand, the resulting Usadel equation needs to preserve a commutator form to ensure the normalization of the quasiclassical GF. 
An alternative and reliable  way to formulate the diffusive limit transport theory
is  via the nonlinear $\sigma$-model approach. 
\cite{wegner1979,efetov1980,finkelshtein1987,kamenev2010,kamenev2011} The saddle-point equation of the non-linear $\sigma$-model is the Usadel equation.  We used this approach previously in Ref.~\onlinecite{virtanen2021} to describe the contribution of extrinsic SOC  to magnetoelectric effects in superconductors. However, in the case of intrinsic SOC described by SU(2) gauge fields, the number of terms appearing in the naive expansion of the action with respect to the gradients and fields is immense and cannot be treated manually.

In this work, we  derive the  nonlinear $\sigma$-model for superconducting and normal  systems  including the intrinsic  SOC in the form of an SU(2) gauge field.  We identify the leading SU(2)
field strength term responsible for the spin Hall effects in the
gradient expansion of the model, and the relevant symmetry
properties. We perform the expansion systematically via computer algebra, and recover also other terms, e.g. describing thermoelectric effects \cite{schwiete2021-nsm}. The
corresponding saddle point equations provide a general framework to
study magnetoelectric effects in disordered superconductors at arbitrary temperature
and generalize the Usadel equation  to include the
intrinsic spin-Hall effects. Using this equation we determine  the anomalous current generated by the interplay between  SOC and an exchange field in a proximitized  normal metal with Rashba SOC.  Our result corrects previous results, and predicts an enhancement of the anomalous current for exchange fields of the order of the superconducting gap. 

The manuscript is structured as follows.  
In Sec.~\ref{sec:mainresults} we outline the main results of our work, namely the 
nonlinear $\sigma$-model Keldysh action, Eqs. (\ref{eq:Sresult}-\ref{eq:S2-baseSH}), and its saddle point equation,  the generalized Usadel equation, Eqs. (\ref{eq:Usadel1}-\ref{eq:Ji1}). 
In Sec.~\ref{sec:grad-exp} we discuss the gradient expansion of the nonlinear $\sigma$-model
with SU(2) gauge fields, and derivation of the main results. Explanation of
the computer implementation of this calculation is postponed to
Appendix~\ref{app:algebra}.  In Sec.~\ref{sec:saddle-point} we derive
the kinetic equations found at the saddle point of the model. In Sec.~\ref{sec:example} we provide an example by calculating the anomalous current induced by an exchange field in a Rashba superconducting system. 
Section~\ref{sec:conclusions} concludes the discussion.

\section{Main results}
\label{sec:mainresults}
Consider a normal conductor with linear-in-momentum spin-orbit coupling (SOC) and an exchange field. In the most  general case, its Hamiltonian can be written as
\begin{equation}
 \label{H-normalSO}
 H_0 = \frac{\hat{\vec{p}}^2 }{2m} - \frac{1}{2m}{\cal A}_{k}^{a}\hat{p}_k \sigma^{a} -\frac{1}{2}{\cal A}_0^a\sigma^{a}+ V_{\rm imp}  \;, 
 \end{equation}
where the second and third  terms describe the SOC and exchange field respectively,  $\sigma^a$ are Pauli matrices spanning the spin space, and $V_{\rm imp}$ is a random impurity potential. Here and throughout the paper summation over repeated indices is implied. 
The linear SOC  can be related  to a local SU(2) gauge invariance of the corresponding Hamiltonian \cite{mineev1992,frohlich1993,jin2006,tokatly2008,gorini2010,bergeret2013,bergeret2014spin}
that can be written (up to an irrelevant constant) as 
\begin{equation}
\label{H-linearSO}
H_0 = \frac{1}{2m}(\hat{p}_j - \hat{\cal A}_j)^2 +V_{\rm imp}-\hat{\cal A}_0, 
\end{equation}
where  $\hat{\cal A}_j=\frac{1}{2}{\cal A}_j^a\sigma^a$. 

To describe superconducting systems with SOC one  constructs from the normal state Hamiltonian, Eq. (\ref{H-linearSO}), the Bogoliubov--de Gennes Hamiltonian
\begin{align}
    \label{eq:Hstart}
    \mathcal{H}
    =
    \tau_3
    \Bigl[
    \frac{[{\hat{\vec{p}}} - \vec{\check{A}}(\vec{r})]^2}{2m} - \mu
    +
    V_{\rm imp}(\vec{r})
    -
    \check{A}_0(\vec{r})
    -
    \hat{\Delta}(\vec{r})
    \Bigr]
    \,,
\end{align}
where $\hat{\Delta}$ is the superconducting anomalous self-energy, for
$s$-wave superconductor given by
$\hat{\Delta}=\tau_3 \tau_1 \Delta e^{-i\tau_3\phi}$.  Here, $\tau_j$
are Pauli matrices in the Nambu  space.
For generality, in Eq.~\eqref{eq:Hstart} we have included the U(1) scalar and vector electromagnetic potentials, $\Phi$, $(A_x,A_y,A_z)$,  
by defining  $\check{A}_i=A_i\tau_3 + \frac{1}{2}\vec{\mathcal A}_{i}\cdot\vec{\sigma}$,
and $\check{A}_0=\Phi + \frac{1}{2}\vec{\mathcal A}_{0}\cdot\vec{\sigma}\tau_3$. 
The field strength associated with $\check{A}$ is
\begin{align}
  \check{F}_{\mu\nu}
  =
  \partial_\mu \check{A}_\nu
  -
  \partial_\nu \check{A}_\mu
  -
  i[\check{A}_\mu, \check{A}_\nu]
  \,,
\end{align}
containing electric and magnetic fields, and their SOC generalizations. Here and below, we use Greek indices for the range $\nu=0,1,2,3$ 
including the time component, and Latin indices for the spatial
components $i=1,2,3$. 

Starting from Hamiltonian \eqref{eq:Hstart}, we derive a disorder-averaged theory describing electron diffusion
in such system,
valid in the quasiclassical diffusive limit
$\xi\gg\ell\gg{}\lambda_F$,
where $\xi$, $\ell$, and $\lambda_F$ are the
superconducting coherence length, the mean free path,
and the Fermi wavelength.
To obtain the Hall and spin-Hall effects, we include the leading sub-quasiclassical corrections, $\propto(\lambda_F/\ell)^1$.
As explained in Sec.~\ref{sec:grad-exp}, we
formulate the problem as a systematic expansion of
Eq.~\eqref{eq:Hstart} in the small parameters, in the approximation
scheme of nonlinear $\sigma$-models, which concentrates on physics
of the low-energy diffusion modes.

\subsection{Keldysh action}

Our main result can be compactly expressed
as the nonlinear $\sigma$-model Keldysh action,
\begin{align}
    \label{eq:Sresult}
    S[Q]
    &=
    S_0[Q]
    +
    S_{\rm H}[Q]
    \,,
    \\
    \label{eq:S2-base}
    S_0[Q]
    &=
    \frac{i\pi\nu_F}{8}
    \Tr
    \Bigl[
    D(\hat{\nabla}Q)^2
    +
    4i(\Omega + \check{A}_0)Q
    \Bigr]
    \,,
    \\
    \label{eq:S2-baseSH}
    S_{\rm H}[Q]
    &=
    \frac{i\pi\nu_FD\ell^2}{8p_F\ell}
    \Tr
    \Bigl[
    -
    \check{F}_{ij} Q (\hat{\nabla}_iQ) (\hat{\nabla}_j Q)
    \Bigr]
    \,.
\end{align}
Here, $Q(\vec{r})$ is a matrix field with  $Q(\vec{r})^2=\mathbb{1}$, which describes
the low energy diffusion modes,
$\hat{\nabla}_iQ=\partial_{r_i}Q - [i\check{A}_i,Q]$ are its covariant
gradients, and $\Omega = i\tau_3\hat{\epsilon} + \hat{\Delta}$
contains the energy operator and local self-energies.
Moreover, $\nu_F$ is the density of states, and $D$ the diffusion constant.

The action $S_0$ comprises the  previous nonlinear $\sigma$-model theory
for superconductivity, \cite{finkelshtein1987,feigelman2000,kamenev2010}
enriched by the spin-dependent gauge fields.

The term  $S_{\rm H}$ contains the leading magnetic field-strength contribution
that breaks the symmetry of $S_0$, hence bringing in
new physical effects related to generalized Hall effects. 
Indeed that term in the case of 2D electron gas and  U(1) fields has the form of the topological term in Pruisken's action for the integer Hall effect \cite{levine1983,pruisken1984-ltq,Altland-book}. We notice that this term also appears in the non-linear sigma model for a disordered Weyl semimetal. \cite{altland2016theory,altland2015effective}
In contrast, the SU(2) counterpart of the Hall term  cannot be written as a total derivative, and hence will generate a nontrivial contribution to the saddle point equation for $Q$ [see Eq.~\eqref{eq:Usadel1}]. $S_{\rm H}$ describes the spin-Hall effects, {\it i.e.} effects caused by the spin charge coupling. 

As discussed later,  the action contains also other terms. For example,  the gradient expansion generates also formally larger terms that do not break symmetries of $S_0$ --- these will renormalize diffusion, but do not contribute to magnetoelectric effects.
There are also other terms  that break 
the electron-hole symmetry, associated with thermoelectric effects, previously 
 discussed in Ref.~\onlinecite{schwiete2021-nsm}. 
We recover all these terms from  our systematic analysis in Sec.~\ref{sec:grad-exp}.

\subsection{Generalized Usadel equation}

The saddle-point equation of $S_0$ is the covariant version of the
well-known Usadel diffusion equation \cite{usadel1970} for
superconducting systems, and Eq.~\eqref{eq:Sresult} provides its
minimal extension including the Hall and intrinsic spin-Hall effects. The Usadel equation is obtained 
from  minimizing the action under the condition $Q^2=1$, that is  $i(\delta S/\delta Q)\rvert_{Q^2=1}=0$, and has 
the form
\begin{align}
   \hat{\nabla}_i \mathcal{J}_i= [i\Omega +i \check{A}_0, Q] 
    \,.
    \label{eq:Usadel1}
\end{align}
Here
$\mathcal{J}_\mu$ are the matrix
currents, proportional to the variation of $S$ with respect to the
matrix-valued vector potential $\check{A}_\mu$. Hence, their
different components are directly related to observable charge $J^c$
and spin $J^s$ currents
by taking appropriate traces, $J^c_i(t)=-\frac{\pi\nu}{2}\tr\tau_3\mathcal{J}_i^K(t,t)$,
$J^s_{ij}(t)=-\frac{\pi\nu}{2}\tr\sigma_j\mathcal{J}_i^K(t,t)$.
Spin density and charge imbalance are given by $S_i=-\frac{\pi\nu}{2}\tr\sigma_i\tau_3J_0^K(t,t)$
and $\delta\rho=-\frac{\pi\nu}{2}\tr{}J_0^K(t,t)$. Here, $K$ superscript denotes the upper right
Keldysh component of the matrix.

The spatial
components of the matrix currents can be expressed as
\begin{align}
  \mathcal{J}_i
  &\equiv
  -
  \frac{2}{\pi\nu}
  \frac{\delta S}{\delta\check{A}_i}
  =
  \mathcal{J}^{(0)}_i 
  +
  \mathcal{J}^{(\mathrm{H})}_i
  \,,
  \\
  \label{eq:J0}
  \mathcal{J}^{(0)}_i 
  &=
  -DQ\hat{\nabla}_iQ
  \,,
  \\
  \label{eq:Ji1}
  \mathcal{J}^{(\mathrm{H})}_i
  &=
  -\frac{D\tau}{4m}
  \Bigl[
  \{\check{F}_{ij} + Q\check{F}_{ij}Q, \hat{\nabla}_jQ\}
  \\\notag&\qquad\qquad
  -
  i\hat{\nabla}_j\bigl(Q[\hat{\nabla}_iQ,\hat{\nabla}_jQ]\bigr)
  \Bigr]
  \,,
\end{align}
and time component is $\mathcal{J}_0^{(0)}=Q$.  
The current $\mathcal{J}^{(0)}$ is the standard diffusive current,
and $\mathcal{J}^{(\mathrm{H})}$ is the leading contribution from spin-charge coupling.
The first term in Eq.~\eqref{eq:Ji1} becomes the (spin-)Hall current in the normal state, and the remainder
gives superconducting corrections.
Eq.~\eqref{eq:Usadel1} implies a covariant conservation equation of these
spin currents \cite{tokatly2008}, where nonconservation of spin current
is associated with the $[-i\mathcal{A},\cdot]$ part of the covariant derivative.

In the next sections we derive the above results.

\section{Gradient expansion}
\label{sec:grad-exp}

The starting point is the Keldysh partition function \cite{kamenev2010,kamenev2011} expressed via the path integral with the action corresponding to the Hamiltonian of Eq.~\eqref{eq:Hstart},
\begin{align}
    S
    &=
    \int_C\dd{t} \bar{\Psi}^T(i\tau_3\partial_t - \tau_3\mathcal{H})\Psi
    \,,
\end{align}
where $\Psi=(\psi_\uparrow,\psi_\downarrow,\bar{\psi}_\downarrow,-\bar{\psi}_\uparrow)^T/\sqrt{2}$ and $\bar{\Psi}=-i\sigma_y\tau_1\Psi$ are Nambu spinors of electron fields on the Keldysh contour $C$.
We then perform standard steps in the nonlinear $\sigma$-model
derivation: (i) averaging the generating function over
the Gaussian disorder potential with $\langle V_{\rm imp}(\vec{r})V_{\rm imp}(\vec{r}')\rangle=\frac{1}{2\pi\nu\tau}\delta(\vec{r}-\vec{r}')$ where
$\tau$ and $\nu$ are  parameters describing the scattering time and density of states, and (ii) decoupling the generated quartic
interaction term with a local matrix field $Q$, $[\bar{\Psi}_i(\vec{r},t)\Psi_{j}(\vec{r},t')][\Psi_i(\vec{r},t)\bar{\Psi}_j(\vec{r},t')]\mapsto\bar{\Psi}_i(\vec{r},t) Q_{ij}(\vec{r};t,t')\Psi_j(\vec{r},t')$. The details of this procedure in the Keldysh formulation are discussed e.g. in Refs.~\onlinecite{kamenev1999,feigelman2000,kamenev2010,kamenev2011}. As these steps only involve the disorder term of the action, the gauge fields do not directly affect the procedure at this stage.

After integrating out the fermion fields,
the result becomes the nonlinear
$\sigma$-model action,
\begin{align}
  \label{eq:S-action}
  S
  &=
  \frac{i\pi\nu_F}{8 \tau}\Tr Q^2
  -
  \frac{i}{2}
  \Tr \ln_\otimes G^{-1}
  \,,
  \\
  G^{-1}
  &=
  \Omega + \mu - \frac{1}{2m}(p_k - \check{A}_k)(p_k - \check{A}_k) + \frac{i}{2\tau}Q
  \,,
\end{align}
where $\Omega = \epsilon\tau_3 + \check{A}_0 + \hat{\Delta}$, 
$\hat{\epsilon} = i\partial_t\delta(t-t')$,
and $\nu_F$ is the density of states at the Fermi energy.

In Keldysh theory, $Q(\vec{r};t,t')$ depends on two times,
and is a $8\times8$ matrix, with $4\times4$ blocks with the Nambu and spin indices,
in a $2\times2$ retarded--advanced--Keldysh structure
\cite{rammer86}.
Here and below, matrix products and trace imply also integrations over time,
$(XY)(t,t')=\int_{-\infty}^\infty\dd{t_1}X(t,t_1)Y(t_1,t')$
and $\Tr X=\int_{-\infty}^\infty\dd{t}\dd{^dr}\tr X(t,t)$.
In some cases it can be technically advantageous to use the energy representation that is defined as follows, 
$X(\epsilon,\epsilon')=\int_{-\infty}^\infty\dd{t}\dd{t'}e^{i\epsilon t-i\epsilon't'}X(t,t')$.

We use here a Wigner representation for the spatial coordinates,
\begin{equation}
  G(\vec{r},\vec{r}';t,t')
  =
  \sum_{\vec{p}} e^{i\vec{p}\cdot(\vec{r} - \vec{r}')} G(\frac{\vec{r}+\vec{r}'}{2}, \vec{p}; t,t')
  \,,
\end{equation}
in which convolutions
$(A\otimes{}B)(\vec{r},\vec{r}';t,t')=\int\dd{\vec{r}_1}\dd{t_1}A(\vec{r},\vec{r}_1;t,t_1)B(\vec{r}_1,\vec{r}';t_1,t')$
can be expressed by the Moyal product
\begin{equation}
  \begin{split}
  (A\otimes{}B)(\vec{r},\vec{p};t,t')
  =
  \int_{-\infty}^\infty\dd{t_1}
  A(\vec{r},\vec{p};t,t_1)
  \\\times
  \exp[\frac{i}{2}(\overset{\leftarrow}{\nabla}_r\cdot\overset{\rightarrow}{\nabla}_p - \overset{\leftarrow}{\nabla}_p\cdot\overset{\rightarrow}{\nabla}_r)]B(\vec{r},\vec{p};t_1,t')
  \,.
  \end{split}
\end{equation}
Here the arrows above the $\nabla$ indicate on which function the derivative operator acts.
The trace becomes
\begin{align}
  \Tr X
  &=
  \int\dd{^dr}\dd{t} \sum_{\vec{p}} \tr X(\vec{r},\vec{p};t,t)
  \,,
\end{align}
where $\tr$ is the trace over matrix indices.  For brevity, in the
following we will not write down the time integrations.

Next, we expand the action of Eq.~\eqref{eq:S-action} in  gradients of $Q$ and
gauge fields $\check{A}_\mu$.
This follows a standard approach in $\sigma$-models.
In this expansion, one usually separates
the ``transverse'' nearly massless modes of the $Q$-field, from the
``longitudinal'' massive modes. The longitudinal modes usually only
renormalize coefficients of the transverse mode theory. 
Below, we concentrate only on the massless modes,
and separate them out by writing $Q=T(\Lambda{}+B)T^{-1}$
where $\Lambda$ is the uniform saddle point solution at $\Omega=\check{A}=0$
with $\Lambda^2=1$,
$B$ with $[B,\Lambda]=0$ describes longitudinal fluctuations, and
$T$ parametrizes the remaining rotations around it. \cite{schafer1980,pruisken1982-ame}
Since at $\Omega=\check{A}=0$ all matrices $Q=T\Lambda{}T^{-1}$ are also saddle-point
solutions, the rotations $T$ in general describe the nearly massless modes, whereas
the longitudinal modes are suppressed by the impurity scattering energy scale $\frac{1}{\tau}$.
We will first neglect the longitudinal corrections and  set $B=0$. 

Inserting the parametrization to Eq.~\eqref{eq:S-action},
the electronic part can be rewritten as
\begin{align}
  \Tr \ln G^{-1}
  &=
  \Tr\ln_\otimes\Bigl[
  \mu - \frac{(p_k-a_k)(p_k-a_k)}{2m} + \frac{ib}{2\tau}\Lambda + a_0
  \Bigr]
  \,,
  \end{align}
  where
  \begin{align}
  a_0 &= T^{-1} (\Omega + \check{A}_0) T
  \,,
  \\
  a_k
  &=
  iT^{-1}\partial_kT + T^{-1} \check{A}_k T
  \,.
\end{align}
The expansion in gradients of $Q$ and in $\check{A}$ and $\Omega$, now
translates to expanding the $\Tr\ln$ in small $a_0$, $a_k$. 
Note that because $\Omega$ contains also the energy $\epsilon$, the expansion 
is valid only at low energies $|\epsilon|\ll{}\tau^{-1}$, and can be used to describe only the
low-energy part of $T(\epsilon,\epsilon')$.

We carry this expansion to
fourth order in $a_i$, $a_0$.  We also expand the result in the
quasiclassical parameter $\psi=p_F\ell\gg1$.  In the end, we rewrite
the result in terms of the covariant gradients
\begin{align}
    &\hat{\nabla}_iQ=T[-ia_i,\Lambda]T^{-1}=\partial_iQ - i[\check{A}_i, Q],\\
    &\hat{\nabla}_0Q=T[-ia_0,\Lambda]T^{-1}=-[i\Omega + i\check{A}_0,Q],
\end{align}
and non-Abelian field strengths
\begin{align} \nonumber
    &T(\partial_ia_j-\partial_ja_i-i[a_i,a_j])T^{-1} \\
    &=\partial_i\check{A}_j - \partial_j\check{A}_i - i[\check{A}_i,\check{A}_j] = \check{F}_{ij},\\
&T(\partial_ka_{0}-i[a_k,a_0])T^{-1} \nonumber \\
&=\partial_k\check{A}_0-\partial_t\check{A}_k-i[\check{A}_k,\check{A}_0] + \hat{\nabla}_k\check{\Delta} = \check{F}_{k0} + \hat{\nabla}_k\check{\Delta}
\end{align}
By construction, the final result should be formally gauge-covariant and therefore it can contain only covariant objects, which is indeed confirmed by explicit calculations.

Expansion of the logarithm, gradient expansion of the Moyal product,
calculation of the momentum sum in the  $\Tr$, and rewriting the result, is  mechanical calculation, and mainly a bookkeeping problem. We discuss our technical method in Appendix~\ref{app:algebra}, and 
concentrate on the results below.

\subsection{Results}

The gradient expansion produces the terms,
\begin{align}
  \label{eq:S-expansion}
  \delta S
  &=
  S_{\rm grad} + S_{\Omega}
  \,,
\end{align}
where $S_{\rm grad}$ contains only spatial gradients, and
$S_{\Omega}$ contains the remaining terms with $\Omega$ and $\check{A}_0$.
The leading part in the expansion of $S_\Omega$ is
well-known,
\begin{align}
    S_\Omega
    &=
    \frac{i\pi\nu_F}{8}
    \Tr 4i(\Omega + \check{A}_0) Q
    +
    \ldots
    \,.
\end{align}
The pure spatial gradient terms, up to fourth order in gradients
and first order in $1/(p_F\ell)$ in three dimensions, are
\begin{align}
    S_{\rm grad}
    &=
    S_2 + S_{4,0} + S_{4,1} + S_{4,1}'
    +
    \ldots
    \,, \label{S-grad}
    \\
    S_2
    &=
    \frac{i\pi\nu_FD}{8}
    \Tr (\hat{\nabla}Q)^2
    \,,
    \\
    \label{S-40}
    S_{4,0}
    &=
    \frac{i\pi\nu_FD\ell^2}{8}
    \Tr\bigl[
    -\frac{3}{5}\hat{\nabla}_{(i}\hat{\nabla}_iQ\hat{\nabla}_j\hat{\nabla}_{j)}Q 
    \\\notag
    &\qquad
    +
    \frac{3}{4}
    \hat{\nabla}_{(i}Q\hat{\nabla}_iQ\hat{\nabla}_jQ\hat{\nabla}_{j)}Q
    \bigr]
    \,,
    \\  
    S_{4,1}
    &\equiv S_H =
    \frac{i\pi\nu_FD\ell^2}{8p_F\ell}
    \Tr\bigl[
    -
    \check{F}_{ij} Q(\hat{\nabla}_iQ)(\hat{\nabla}_j Q)
    \bigr]
    \,, \label{S-41}
    \\
    S_{4,1}'
    &=
    \frac{i\pi\nu_FD\ell^2}{8p_F\ell}
    \Tr\bigl[
    iQ(\hat{\nabla}_i\hat{\nabla}_jQ)(\hat{\nabla}_i\hat{\nabla}_jQ)
    \bigr]
    \,. \label{S'-41}
\end{align}
The double expansion consists of terms $S_{m,n}$ that are of order $m$ in gradients,
and order $n$ in $1/(p_F\ell)$. Here, braces in indices denote symmetrization of the tensor, 
which is defined as the average $X_{(i_1,\ldots,i_N)}=\frac{1}{|\mathcal{P}|}\sum_{\sigma\in\mathcal{P}} X_{i_{\sigma(1)},\ldots,i_{\sigma(N)}}$
over all permutations $\sigma$, where $\mathcal{P}$ is the set of permutations of $1,\ldots,N$ and $|\mathcal{P}|=N!$.

The prefactor of the diffusion term $S_2$ is related to the
longitudinal Drude conductivity by
$\nu_FD=e^{-2}\sigma_{xx}$. Similarly, the prefactor of the field
strength term $S_{4,1}=S_H$ is
related to the transverse Hall conductivity
by $\nu_FD\ell^2/(p_F\ell)=e^{-3}\frac{d\sigma_{yx}}{dB}\rvert_{B=0}$
where $\sigma_{yx}=\sigma_{xx}\omega_c\tau$ and $\omega_c=\frac{eB}{m}$ is the cyclotron frequency.  By the
generic gauge structure of the theory, this coefficient is the {\it
  same} both for the Hall effect and the spin-Hall effect.  One can
also note that unlike in the quantum Hall effect
\cite{levine1983,pruisken1984-ltq,Altland-book}, we are here working in the limit of
small field strengths, and so the prefactor is not quantized.

While such physical considerations suggest that the relationship
between the coefficients of these two terms is fixed, in general the
coefficients of the higher-order gradient terms can be modified by the
longitudinal fluctuations of the $Q$ field.  Already if we allow for
nonzero longitudinal part $B\ne0$ on the saddle-point level,
\cite{supplement} additional corrections of similar order in the small
parameters as in $S_{4,0}$ and $S_{4,1}'$ appear and additional
considerations are necessary if one wants to derive coefficients of
such terms from the microscopic theory. On the saddle-point level, the
coefficient of $S_{4,1}$ is not renormalized by them.

The higher-order gradient terms in Eq.~\eqref{eq:S-expansion} are
usually not of direct physical interest (see, however,
\cite{kravtsov1988-wlt,evers2008-at}), except if they break a
symmetry present in the lower-order model $S_2$. In that case, they can
give rise to new physically interesting phenomena.  We discuss the
model symmetry in more detail in Sec.~\ref{sec:symmetries}.

We can recognize that the non-Abelian field strength term
$S_{4,1}=S_H$ is a generalization of the spin-Hall term in
Ref.~\onlinecite{virtanen2021} from the extrinsic spin-Hall effect to
the intrinsic one, with the correspondence
$\check{F}_{ij}\leftrightarrow-\frac{p_F}{\ell}\theta\epsilon_{ijk}\sigma_k$,
where $\theta$ is the spin-Hall angle.  In the equations of motion,
this term will lead to conversion between charge  and spin
currents, a feature that is not present in a model containing only
$S_2$. The term in $S_{4,1}'$ on the other hand was previously noted
in Ref.~\onlinecite{schwiete2021-nsm} in the context of  thermoelectric
effects for $\vec{A}=F=0$.

\subsection{Symmetry analysis}
\label{sec:symmetries}

Terms in the action $S$ can be classified based on which
symmetries they break.
A symmetry relevant for magnetoelectric effects is
the electron-hole symmetry, or the "quasiclassical" symmetry,
which implies \cite{silaev2017} a certain lack of coupling
between spin and charge transport.
The electron-hole symmetry can be understood as one that flips the
sign of the electron dispersion,
$\xi_p=\frac{p^2}{2m}-\mu\mapsto-\xi_p$.  On the level of the
$\sigma$-model, this can be expressed as the transformation \cite{silaev2017},
\begin{align} \label{e-h-symmetry}
  Q
  \mapsto
  \tilde{Q}
  \equiv
  -\tau_1 \bar{Q} \tau_1
  =
  -\tau_2\sigma_y Q^T \sigma_y \tau_2
  \,,
\end{align}
where $\bar{Q}=\tau_3\sigma_y Q^T \sigma_y\tau_3$ is a time-reversal
transformation.  Similarly as for Green functions, the transformation
swaps the Nambu blocks and reverses time, which corresponds to
reversing sign of the "normal" part in the Bogoliubov--de Gennes Hamiltonian while keeping the anomalous unchanged.  A similar
transformation was used in Ref.~\onlinecite{schwiete2021-nsm}; the
above is its extension to superconducting case, which requires keeping also the superconducting anomalous self-energy terms
$\propto \Tr[\Delta\tau_3\tau_1e^{i\tau_3\varphi}Q]$
invariant. 

In the usual $\sigma$-model the action $S_0$ consists of  $S_2$ and the leading term in $S_{\Omega}$.  In the latter term, $\check{A}_0$ changes sign under the e-h transformation Eq.~\eqref{e-h-symmetry}, whereas $S_2$ remains invariant. This in particular implies that at the level of $S_0$, the transformation Eq.~\eqref{e-h-symmetry} defines a mapping between systems with opposite directions of the Zeeman field $\vec{\mathcal A}_{0}$. When combined with the time reversal symmetry, this mapping forbids the anomalous supercurrent \cite{silaev2017,Rasmussen2016}, which explains  the absence of magnetoelectric effects in the theory defined by the leading contribution $S_0$ to the $\sigma$-model. 

By analyzing the subleading higher gradient terms in Eq.~\eqref{S-grad}, we find that $S_{4,0}$ is invariant under the transformation of Eq.~\eqref{e-h-symmetry}, but $S_{4,1}=S_H$ and $S_{4,1}'$ change sign. Expansion of $S_\Omega$ (see Appendix~\ref{app:results}) can be classified similarly.
Hence, even though $S_{4,0}$ is formally larger in the quasiclassical parameter $p_F\ell\gg1$, 
it has the same symmetry as $S_2$, and we expect its effect is merely a renormalization of diffusion and not of interest for us here. In contrast, the terms breaking the "quasiclassical" e-h symmetry introduce new physics and qualitatively change the behaviour of the system.

The first antisymmetric term $S_{4,1}=S_H$, Eq.~\eqref{S-41}, introduces the spin Hall effect. This term is responsible for all magneto-electric effects mediated by intrinsic SOC, such as direct and inverse spin-galvanic/Edelstein effects in normal conductors and the appearance of anomalous supercurrents and anomalous Josephson effect in superconductors.  

The second antisymmetric term,
$S_{4,1}'$ of Eq.~\eqref{S'-41}, and the term $\frac{i\pi\nu\ell}{8p_F}\Tr \Omega(\hat{\nabla}_iQ)(\hat{\nabla}_iQ)$ appearing
in $S_{\Omega}$ (see Appendix~\ref{app:results})
were previously presented in Ref.~\onlinecite{schwiete2021-nsm} providing  an extension of the $\sigma$-model  to include thermoelectric effects.

\section{Saddle point}
\label{sec:saddle-point}

As noted in Sec.~\ref{sec:mainresults},
variation of the action produces the generalized Usadel
equation that is the saddle point condition for $S[Q]$.

Technically, the variation under the condition
$Q^2=1$ is calculated by writing $Q=T\Lambda T^{-1}$, where $\Lambda$ is the uniform saddle point solution, and by observing  that  $\delta Q = [W,Q]$ with $W=(\delta T)T^{-1}$.
A straightforward calculation \cite{supplement} with subsequent integration
by parts then produces the final result, which can however be represented in different forms, reflecting different formal properties of the saddle point equation. On the one hand, the commutator form of the allowed variations $\delta Q = [W,Q]$ implies that the saddle point equation also has a commutator form $[\dots,Q]=0$, which guarantees that it is consistent with the normalization condition $Q^2=1$. On the other hand, the gauge invariance of the action implies that the saddle point equation can be represented in a form of a covariant conservation law,
\begin{align} \label{Usadel}
   \hat{\nabla}_i \mathcal{J}_i= [i\Omega +i \check{A}_0, Q] 
    \,,
\end{align}
where the matrix current $\mathcal{J}_i$ can be expressed as variational derivative of the action with respect to the gauge potential
\begin{align}
  \mathcal{J}_i
  &\equiv
  -
  \frac{2}{\pi\nu}
  \frac{\delta S}{\delta\check{A}_i}
  =
  \mathcal{J}^{(2)}_i 
  +
  \mathcal{J}^{(4,0)}_i
  +
  \mathcal{J}^{(4,1)}_i
  +
  \mathcal{J}^{(4,1')}_i
  \,,
\end{align}
where from $S_{2}$ and $S_{4,1}$ we obtain
\begin{align}
  \label{eq:J2}
  \mathcal{J}^{(2)}_i 
  &=
  -DQ\hat{\nabla}_iQ
  \,,
  \\ 
  \label{eq:J41}
  \mathcal{J}^{(4,1)}_i
  &=
  -\frac{D\tau}{4m}
  \Bigl[
  \{\check{F}_{ij} + Q\check{F}_{ij}Q, \hat{\nabla}_jQ\}
  \\\notag&\qquad\qquad
  -
  i\hat{\nabla}_j\bigl(Q[\hat{\nabla}_iQ,\hat{\nabla}_jQ]\bigr)
  \Bigr]
  \,.
\end{align}
Similarly, from $S_{4,0}$ and $S_{4,1}'$:
\begin{align}
    \mathcal{J}_i^{(4,0)}
    &=
    -\frac{3D\ell^2}{5}\Bigl(
    Q\hat{\nabla}_{(i}\hat{\nabla}_{j}\hat{\nabla}_{j)}Q
    \\\notag&\quad
    + Q[\hat{\nabla}_{(i}Q\hat{\nabla}_{j}\hat{\nabla}_{j)}Q + 2\hat{\nabla}_{(i} \hat{\nabla}_{j}Q\hat{\nabla}_{j)}Q]Q
    \\\notag&\quad
    + \frac{1}{2}Q\hat{\nabla}_{(i}Q\hat{\nabla}_{j}Q\hat{\nabla}_{j)}Q
    \Bigr)
    \,,
    \\
    \mathcal{J}_i^{(4,1')}
    &=
    -\frac{D\tau}{2m}
    [\hat{\nabla}_iQ, \hat{\nabla}_jQ\hat{\nabla}_jQ]\label{j41p}
    \,.
\end{align}
As we noticed before, whereas the numerical pre-factors in Eqs. (\ref{eq:J2}-\ref{eq:J41}) are not modified on the saddle-point level \cite{supplement} by the longitudinal corrections, pre-factors in the last two equations have to be renormalized.

At this point it is instructive to compare the present covariant theory with intrinsic SOC, and the theory of diffusive systems with SOC of extrinsic origin, such as random impurities \cite{virtanen2021}. At the level of the nonlinear $\sigma$-model, the extrinsic and intrinsic theories can be connected by (i) replacing the usual gradients with the covariant gradients, and (ii) identifying $\check{F}_{ij}\leftrightarrow-\frac{p_F}{\ell}\theta\epsilon_{ijk}\sigma_k$ in the spin Hall term. It is natural to expect that the same replacement rules should work for the Usadel equation. However, this does not look obvious if one naively compares the Usadel equation from Ref.~\cite{virtanen2021} and Eq.~\eqref{Usadel} with the current given by Eqs.~\eqref{eq:J2} and \eqref{eq:J41}. In the extrinsic case an additional torque term $\mathcal{T}$ appears in the Usadel equation, while the current has a form that can be identified only with the first term in Eq.~\eqref{eq:J41}. This apparent contradiction is resolved by noticing that the expected torque is of course there, but because of the gauge symmetry it must have a form of a covariant divergence. It therefore appears in the Usadel equation as an additional contribution to the current. Indeed, by using the identity $-i[\check{F}_{ij},f]=[\hat{\nabla}_i,\hat{\nabla}_j]f$ and the replacement $\check{F}_{ij}\mapsto-\frac{p_F}{\ell}\theta\epsilon_{ijk}\sigma_k$, we recover the correct torque $\mathcal{T}$ of the extrinsic theory. Also, in the case of extrinsic SOC, there is a  spin relaxation Elliot-Yafet term,  in the Usadel equation. In the present case of Eq.~\eqref{Usadel}, the spin relaxation stems from the double covariant gradient term when substituting Eq. (\ref{eq:J2}) into Eq.~(\ref{Usadel}). This term corresponds to the Dyakonov-Perel type of spin relaxation.

Finally we make a connection to the well established theory of normal conductors with SOC. The normal-state diffusion equations can be recovered from Eq.~\eqref{Usadel} 
by the replacement
\begin{align}
  Q \mapsto
  \begin{pmatrix}
    1 & 2 f
    \\
    0 & -1
  \end{pmatrix}
  \,,
\end{align}
where the $2\times2$ matrix is in Keldysh space, and Nambu
structure is trivial and we replace $\tau_3\mapsto1$.  Here,
$f(\vec{r},t,t')=f_0(\vec{r},t,t')+\vec{\sigma}\cdot\vec{f}(\vec{r},t,t')$ is the spin-dependent
distribution function of electrons.
The Usadel equation then becomes
the diffusion equation
\begin{align}
  \label{eq:UsadelN}
  [i\hat{\epsilon}, f]
  &=
  \hat{\nabla}_i\mathcal{J}_i
  \,,
  \\
  \label{eq:UsadelNJ}
  \mathcal{J}_i
  &=
  -
  D \hat{\nabla}_if
  -
  \frac{D\tau}{2m}\{\check{F}_{ij}, \hat{\nabla}_j f\}
  -
  \frac{3D\ell^2}{5}\hat{\nabla}_{(i}\hat{\nabla}_j\hat{\nabla}_{j)} f
  \,,
\end{align}
where we wrote the terms corresponding to the leading term of
$S_\Omega$, and the spatial gradient terms $S_2$, $S_{4,1}$, and also
$S_{4,0}$ which gives a symmetrized derivative term.  The other
higher-gradient term $S_{4,1}'$ gives no contribution in normal state.

Keeping only the first two terms in Eq.~\eqref{eq:UsadelNJ}, when
integrated over the energy, results to the known spin-charge diffusion
equations in the case of normal metals with intrinsic
SOC. \cite{burkov2004,shen2014-tcs,gorini2010,raimondi2012spin,sanz2019nonlocal} In the case of $U(1)$ magnetic field strength $\vec{B}$, one can derive  the equations describing the ordinary  Hall
effect. 

The last symmetric derivative term in Eq.~\eqref{eq:UsadelNJ} from
$S_{4,0}$ is essentially always neglected in derivations of such
normal-state equations, even though it is formally larger by
$p_F\ell\gg1$ than the SOC term. As we argued in
Sec.~\ref{sec:symmetries}, it can however be excluded on symmetry
grounds.  For completeness, let us show how this term would appear in
standard derivations.  In the normal state, we can consider the
quasiclassical distribution function $f(\hat{p},\vec{r};t,t')$ on the
Fermi level, which depends on the position and momentum direction.  It
obeys a transport (Eilenberger) equation \cite{eilenberger1968-tog}
\begin{align}
  -\frac{1}{\tau}(f - \avg{f}) = (\vec{v}\cdot\hat{\nabla} + \partial_t - \partial_{t'})f = \hat{D}f
  \,,
  \label{eq:eilenbergerN}
\end{align}
where $\hat{\nabla}$ is the gauge-invariant gradient and
$\vec{v}(\hat{p})$ the velocity.
This formulation works on the $1/(p_F\ell)^0$ level, and will not capture magnetoelectric effects.
We can formally solve Eq.~\eqref{eq:eilenbergerN}, take the average $\langle\cdot\rangle$
over momentum directions $\hat{p}$, and expand in $\tau\to0$ to find
\begin{align}
  \label{eq:f-expansion}
  0
  &=
  \frac{1}{\tau}
  \langle
  (1 + \tau \hat{D})^{-1} \langle f \rangle
  -
  f\rangle
  =
  \frac{1}{\tau}\sum_{n=1}^\infty \avg{[-\tau\hat{D}]^n} \avg{f}
  \,.
\end{align}
Truncating to fourth order in spatial gradients and first order in
time derivative, this gradient expansion becomes
\begin{align}
  0
  &=
  -(\partial_t-\partial_{t'}) \langle f \rangle
  +
  \tau\langle{v_iv_j}\rangle \hat{\nabla}_i\hat{\nabla}_j \langle f \rangle
  \\\notag&\qquad
  +
  \tau^3\langle{v_iv_jv_kv_l}\rangle \hat{\nabla}_i\hat{\nabla}_j\hat{\nabla}_k\hat{\nabla}_l \langle f \rangle
  \\
  &=
  -(\partial_t-\partial_{t'}) \langle f \rangle
  +
  D \hat{\nabla}_i\hat{\nabla}_i \langle f \rangle
  +
  \frac{D \ell^2 d}{d + 2} \hat{\nabla}_{(i}\hat{\nabla}_i\hat{\nabla}_j\hat{\nabla}_{j)} \langle f \rangle
  \,,
\end{align}
where $d$ is the space dimension. The first two terms constitute the
standard diffusion equation, and the third term is what appears in
Eq.~\eqref{eq:UsadelN}, recognizing that
$\hat{\nabla}_i\hat{\nabla}_{(i}\hat{\nabla}_{j}\hat{\nabla}_{j)}=\hat{\nabla}_{(i}\hat{\nabla}_i\hat{\nabla}_j\hat{\nabla}_{j)}$.

\section{Example: Anomalous current in a superconducting system with Rashba  SOC}\label{sec:example}

It was predicted first by Edelstein \cite{edelstein1995,edelstein2005magnetoelectric} that a  superconductor with a Rashba SOC supports spontaneous supercurrents in the presence of a Zeeman field. This is nothing but the superconducting version of the spin-galvanic effect predicted in normal systems \cite{shen2014}. Anomalous supercurrents are not only present in superconductors but also in proximitized normal systems and  Josephson junctions with SOC \cite{buzdin2008direct,bergeret2015theory,konschelle2015theory,strambini2020josephson,assouline2019spin}.  

The spin-galvanic effect in superconductors has been studied primarily on linear response assuming either a small superconducting gap $\Delta$ or a small Zeeman field ${\cal A}_0$ \cite{edelstein2005magnetoelectric, buzdin2008direct,bergeret2015theory,konschelle2015theory,strambini2020josephson,assouline2019spin,bergeret2020theory}.    Going beyond linear response 
is not straightforward.  The result of  Ref.~\onlinecite{tokatly2017} suggested that the spin--galvanic relation between the charge current and the induced spin in a superconductor is identical to that in a normal metal, {\it i.e.} that the current induced by the SOC is proportional to the deviation of the spin density from the Pauli response, $\delta S$. 
However, as follows from Eq.~(\ref{eq:Ji1}),   this statement is incorrect.  In this section, we determine the anomalous current induced in a Rashba superconductor to all orders in the magnetic field and  arbitrary temperatures.

Let us consider a two-dimensional infinite homogeneous normal system with an isotropic Rashba SOC, proximitized by a superconductor.  The SOC is described by the SU(2) vector potential with components $\mathcal{A}_x=2\alpha \sigma_y$ and  $\mathcal{A}_y=-2\alpha \sigma_x$. Because the system is homogeneous, no gradient terms enter Eq.~(\ref{Usadel}). Moreover, the term Eq.~(\ref{eq:J41}) does not contribute to the Usadel equation which acquires the simple form\cite{tokatly2017}:
\begin{align} \label{UsadelRashba}
   D\alpha^2 \left[\sigma_i,Q \left[\sigma_i,Q\right]\right]= [i\Omega +i h\sigma_x, Q] 
    \,.
\end{align}
The  term in the left-hand side is the Dyakonov-Perel relaxation term
due to the SOC, and we have assumed that the Zeeman field, $h$, points in  $x$-direction. 
From this equation one determines the function $Q$ which in the present situation has the structure \begin{equation}
Q=\hat Q_0+\hat Q_x \sigma_x\; ,  \label{eq:fullQ}    
 \end{equation}
 where $\hat Q_{0,x}$ are 2$\times$2 matrices in Nambu space.

Once the matrix $Q$ is obtained one can determine the Hall matrix current, $J^{(H)}$,  given by Eq.~\eqref{eq:Ji1}:
\begin{equation}
    J_{an}=-\frac{D\tau}{4m} (i\pi T\nu)\sum _{\omega_n}{\rm Tr \tau_3}\{\check{F}_{yx}+Q\check{F}_{yx}Q,\hat\nabla_x Q\}\; . \label{eq:ex_current}
\end{equation}
Notice that the last term in Eq.~\eqref{eq:Ji1} does not contribute in the case considered here.
This expression  has the same structure of the matrix current derived in Ref.~\onlinecite{tokatly2017} after the  substitution $\hat{\mathcal{F}}_{ij}\rightarrow (\hat{\mathcal{F}}_{ij}+Q\hat{\mathcal{F}}_{ij}Q)/2$. As we demonstrate next the term $Q\hat{\mathcal{F}}_{ij}Q$ leads to a enhancement of the anomalous current when $h\sim\Delta$. 
 
First,  in order to obtain an analytical result we  neglect the relaxation term and obtain for the components of  $Q$ in Eq.~(\ref{eq:fullQ})
 \begin{equation}
     \hat Q_{0,x}=g_{0,x}\tau_3+f_{0,x}\tau_1 
 \end{equation}
 with 
 \begin{eqnarray}
 g_{0,x}&=&\frac{g_+\pm g_-}{2}\\
  f_{0,x}&=&\frac{f_+\pm f_-}{2}\; , 
 \end{eqnarray}
 and 
 \begin{eqnarray}
 g_\pm&=&\frac{\omega_n\pm i h}{\sqrt{(\omega_n\pm i h)^2+\Delta^2}}\\
 f_\pm&=&\frac{\Delta}{\sqrt{(\omega_n\pm i h)^2+\Delta^2}}\label{eq:fpm} \; .
 \end{eqnarray}
 We have used the Matsubara representation of the Green's functions , with $\omega_n=\pi T (2n+1)$, and $h$ is the amplitude of the Zeeman or exchange field. 
 
 It is easy to check that the $Q$ defined by Eqs.~(\ref{eq:fullQ}-\ref{eq:fpm}) satisfied the normalization $Q^2=1$.  To calculate the charge anomalous current, $J_{an}$ one substitutes the above $Q$ in   Eq.~\eqref{eq:ex_current}:
\begin{eqnarray}
    J_{an}&=&-\frac{i\pi T D\tau\alpha^3}{2m}\sum_{\omega_n}(g_+-g_-)(1+g_+g_-+f_+f_-)  \label{eq:final_Jan}\\ &=&\frac{2\pi T D\tau\alpha^3}{m}h\Delta^2\sum_{\omega_n}\frac{{\rm Re}\sqrt{\Delta^2+(\omega_n+i h)^2}}{h^4+2h^2(\omega_n^2-\Delta^2)+(\omega_n^2+\Delta^2)^2}\nonumber
\end{eqnarray} 
The value of the maximum of $J_{an}$ at $h=\Delta$ in the limit $T\ll\Delta$  can be found as follows. For small $T$ and $h=\Delta$ the Matsubara sum in Eq.~\eqref{eq:final_Jan} is dominated by $\omega\ll\Delta$. Therefore,
\begin{align} \nonumber
    J_{an}\vert_{h=\Delta} &\approx \frac{\pi\Delta D\tau\alpha^3}{m}
    T\sum_{\omega >0}\frac{\sqrt{\Delta}}{\omega^{3/2}}\\
    &=\frac{\pi\Delta D\tau\alpha^3}{m}\frac{2^{3/2}-1}{(2\pi)^{3/2}}
    \zeta(3/2)\sqrt{\frac{\Delta}{T}}
\end{align}
This result holds when  neglecting the spin-relaxation. If the latter is taken into account, 
the $1/\sqrt{T}$ divergence at $T\to 0$ saturates to the value $1/\sqrt{\Gamma_{DP}}$ at temperatures smaller than the Dyakonov-Perel spin relaxation rate $\Gamma_{DP}=D\alpha^2$.
    
\begin{figure}[h!]
\includegraphics[width=0.48\textwidth]{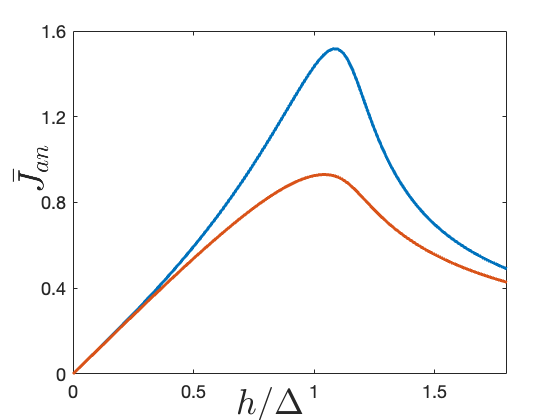}
\caption{The anomalous current $\bar J_{an}=J_{an} 2m/(\pi\Delta D\tau\alpha^3)$ (blue solid line) as a function of the field $h$ for $D\alpha^2=0.1\Delta$ and $T=0.02 \Delta$. The red line   shows the result obtained from  Ref.~\onlinecite{tokatly2017}.\label{Figure}}
\end{figure}

Inclusion of the spin-relaxation term stemming after substitution of  Eq.~(\ref{eq:Ji1}) into Eq.~(\ref{eq:Usadel1} does not allow for a analytical solution for the Green's functions. We therefore  calculate numerically the anomalous current, Eq.~(\ref{eq:final_Jan}).      In Fig. \ref{Figure} we show the result for $D\alpha^2=0.1\Delta$ at $T=0.5\Delta$. The red  line shows the result obtained using the expression for the current from Ref.~\onlinecite{tokatly2017}, $J_{an}=-i\pi T D\tau\alpha^3/(2m)\sum_{\omega_n}(g_+-g_-)$. Whereas at small field both results coincide (linear regime), the anomalous current at fields comparable to $\Delta$ is clearly larger than the one  obtained in that work.

\section{Conclusions}
\label{sec:conclusions}

With the help of computer algebra, we have derived the nonlinear $\sigma$-model for superconducting and normal systems,  including the intrinsic  SOC. The latter enters the theory as an effective SU(2) gauge field and therefore appears in the $\sigma$-model only via gauge covariant combinations, the field strength, and covariant derivatives. We have performed a systematic gradient expansion and identified the spin Hall term, which is responsible for the spin Hall effect and other magnetoelectric effects mediated by the intrinsic SOC, such as spin-galvanic/Edelstein effects and anomalous supercurrents in superconductors. In the same order of the gradient expansion, we also recover the previously discussed contribution related to thermoelectricity \cite{schwiete2021-nsm}.

The saddle point equation of the model, Eqs.~(\ref{Usadel}-\ref{j41p}), which corresponds to the generalized Usadel equation, reveals new terms only present in the superconducting state and in nonlinear regimes, e.g. Eq.~(\ref{eq:J41}). We applied the derived equations to compute the anomalous current generated by a Zeeman field in a superconductor with Rashba SOC. We observe a substantial increase in the anomalous current compared to the results of previous incomplete theories, which clearly demonstrates the importance of new nonlinear terms in the saddle point equations.

The presented generalized nonlinear $\sigma$-model provides a flexible and convenient tool for studying diffusive superconducting systems with intrinsic SOC. It is expected to be especially useful for analyzing the effects of intrinsic spin-charge coupling in the nonlinear regime at the saddle point level and for the description of a wide range of fluctuation phenomena.

\acknowledgments

We thank R. Raimondi for useful discussions and for his input on the early stage of this work.

P.V. and F.S.B. acknowledge funding from EU's Horizon 2020 research and
innovation program under Grant Agreement No. 800923 (SUPERTED). P.V.
acknowledges funding from Academy of Finland Project 317118. I.V.T.
acknowledges support by Grupos Consolidados UPV/EHU del Gobierno Vasco (Grant
No. IT1249-19). F.S.B.  and I.V.T acknowledge funding by the Spanish Ministerio de
Ciencia e Innovacion  (MICINN) (Projects PID2020-117671GB-I00  (SPIRIT) and PID2020-112811GB-I00). F.S.B thanks Prof. Bj\"orn Trauzettel for his hospitality at W\"urzburg University, and  the A. v. Humboldt Foundation for financial support. 

\bibliography{nlsmsoc}

\appendix

\section{Computer algebra methods}

\label{app:algebra}

\subsection{Gradient expansion}

We compute the expansion of
\begin{align}
  \label{eq:delta-S-exp}
  \delta S
  &=
  -\frac{i}{2}
  \Tr\ln G^{-1}
  +
  \frac{i}{2}
  \Tr\ln \mathcal{G}^{-1}
  \notag
  \\
  &=
  -\frac{i}{2}
  \Tr\ln_\otimes\Bigl[1 + \Bigl(\frac{2p_ka_k - a_ka_k}{2m} + a_0\Bigr)\otimes\mathcal{G}\Bigr]
  \,,
\end{align}
in the small parameter $a_k,\partial_{k}\sim\mathrm{small}$.  Here,
$\mathcal{G}_{\vec{p}}^{-1}=\mu-\frac{p^2}{2m} +
\frac{i}{2\tau}\Lambda$, and $\Lambda^2=1$.

As explained in the main text, we work in the Wigner representation, where
the Moyal product has the gradient expansion
\begin{align}
  X\otimes{}Y
  &=
  \sum_{n=0}^\infty
  \frac{i^n}{2^nn!}
  X \Bigl(\overset{\leftarrow}{\nabla}_r\cdot\overset{\rightarrow}{\nabla}_p - \overset{\leftarrow}{\nabla}_p\cdot\overset{\rightarrow}{\nabla}_r\Bigr)^n Y
  \,,
\end{align}
that corresponds to an expansion in ``small'', and can be truncated.
It is convenient to express the mass $m=\tau\psi/\ell^2$ and the
chemical potential $\mu=\psi/(2\tau)$ in terms of the scattering time
$\tau$ or mean free path $\ell$, and the quasiclassical parameter
$\psi=p_F\ell$, to express the result as a double expansion in small
$\ell$ and large $\psi$.

The series expansion of the logarithm and the Moyal product, and (as
we see below) calculation of the momentum sum is straightforward.  It
results to the gradient expanded action, which can in the end be
re-expressed in terms of gauge-invariant derivatives
$\hat{\nabla}_jQ=\partial_{j}Q-[iA_j,Q]$ and the field strength
$\check{F}_{ij}=\partial_iA_i - \partial_jA_j - i[A_i,A_j]$. We truncate the
expansion in order $\mathrm{small}^4$, which is where the lowest-order
spatial field-strength term appears.

As the calculation of many ($>100$) terms is tedious to do manually, we implement this in
computer algebra.  The gradient expansion outlined above can be done
symbolically by considering an algebra of terms $c m$ consisting of a
noncommutative monomial $m$ and a scalar coefficient $c$, with $m$ consisting
of a product of symbols $\{\Lambda, a_i, a_0, \mathcal{G}\}$.  The
only nonzero momentum derivative of the base symbols is
$\partial_{p_i}\mathcal{G}=\frac{p_i}{m}\mathcal{G}\mathcal{G}$.  The
nonzero spatial derivatives we denote in terms of
$a_{X,i}\equiv\partial_ia_X$ where $X$ is some set of indices. Here,
$a_{ijk\ldots}$ are symmetric under exchange of indices, excluding the
first index.  The order in $\mathrm{small}$ of a monomial is equal to
the total number of indices in $a$.

At the end of the expansion, the monomials appear under $\tr$, where
they can be permuted cyclically.  We define an (arbitrary) ordering of
monomials $m\succ{}m'$, and permute each term to the order where the
monomial is minimal. The procedure produces an expansion
$\delta S = \Tr \sum_j c_jm_j$ which after momentum integration
becomes the local gradient expansion
$\delta S= \int\dd{^dr} \tr \sum_j c_j' m_j'$.  We will also drop
total derivative terms.  Finally, we re-express $\delta S$ in a gauge-invariant form,
in terms of monomials of symbols $Q_{X,i}\leftrightarrow{}\hat{\nabla}_iQ_X$ and
$\check{F}_{ij}$.

It is also possible to formulate the expansion in a manifestly
gauge-invariant manner, in terms of a gauge-invariant Moyal
product. However, the present approach is simpler to implement in
computer algebra.

The implementation is written using {\it SageMath} \cite{sagemath94},
and is included in the Supplementary Material \cite{supplement}.

\subsection{Momentum integration}
\label{app:momentum-int}

The momentum integrals are calculated exactly, and then expanded in
series in the quasiclassical parameter $\psi=p_F\ell=2\mu\tau\gg1$.
Although their analytic evaluation is standard, for completeness, we
explain it here in a form suitable for straightforward computer
implementation.

All expressions generated by the expansion have the form
\begin{align}
  I
  &= \sum_{\vec{p}} f(\vec{p})
  = \int_{-\mu}^\infty\dd{\xi}\nu(\xi) \langle{}f(\vec{p}(\xi))\rangle
  \,,
  \\
  f(\vec{p}) &= g(\vec{p}) \mathcal{G}_{\vec{p}} Z_1 \mathcal{G}_{\vec{p}} Z_2 \ldots Z_{N-1} \mathcal{G}_{\vec{p}}
  \,,
\end{align}
where $\mathcal{G}_{\vec{p}}^{-1}=-\xi_{\vec{p}} + \frac{i}{2\tau}\Lambda$,
and we consider a parabolic band $\xi_{\vec{p}}=\frac{p^2}{2m}-\mu$
in $d$ dimensions, $\nu(\xi)=(1 + \xi/\mu)^{d/2-1}\nu_F$.
Here, $g(\vec{p})$ is a product of $p_i$ and a $\vec{p}$-independent scalar,
and $Z_j$ are momentum-independent.

The angular average $\langle\cdot\rangle$ over the $(d-1)$-sphere can be
evaluated with a well-known formula, for $n$ even,
\begin{align}
  \label{eq:dsphere-avg}
  \int_{S^{d-1}}\frac{\dd{S_p}
    p_{i_1} \ldots p_{i_n}
  }{V(S^{d-1})}
  =
  \frac{p^n(d - 2)!!}{(n + d - 2)!!}
  \sum_{C\in\mathcal{P}}\prod_{(i_a,i_b)\in{}C}\delta_{i_a,i_b}
  \,,
\end{align}
where $\mathcal{P}$ is the set of all pairings $C$ of the indices $\{i_1,\ldots,i_n\}$,
and $V(S^d)$ is the sphere surface volume. The result is zero if $n$ is odd.
We then get $\langle{}g(\vec{p})\rangle=g_0 \bigl(\frac{p^2}{2m}\bigr)^\beta$,
where $\beta$ is an integer and $g_0$ a scalar.

Using $\Lambda^2=1$ and that
$\mathcal{G}_{\vec{p}}=(-\xi_{\vec{p}}-\frac{i}{2\tau}\Lambda)/(\xi_{\vec{p}}^2+\frac{1}{4\tau^2})$,
we can rewrite
\begin{align}
  \label{eq:xi-integration-formula}
  I
  &=
  -i
  \pi
  \nu_F
  \mu^{\beta+1-N}
  g_0
  \sum_{\vec{\alpha}}
  C_{\beta,N,|\vec{\alpha}|_1} R_{\vec{\alpha}}
  \,,
  \\
  R_{\vec{\alpha}}
  &=
  \Lambda^{\alpha_1} Z_1 \Lambda^{\alpha_2} Z_2 \ldots Z_{N-1} \Lambda^{\alpha_N}
  \,,
\end{align}
where the sum runs over the multi-index
$\vec{\alpha}=(\alpha_1,\ldots,\alpha_N)$, $\alpha_j\in\{0,1\}$, and
$|\vec{\alpha}|_1=\sum_j\alpha_j$.  The coefficients are given by:
\begin{align}
  C_{\beta{}N{}n}
  &=
  \frac{
    i^{n+1}
    (-1)^N
  }{\pi\psi^n}
  \int_{-1}^\infty\dd{z}
  \frac{
    (1 + z)^{\beta + d/2 - 1} z^{N-n}
  }{
    (z^2 + \psi^{-2})^N
  }
  \,,
\end{align}
where $\psi=2\mu\tau=p_F\ell$.

\begin{table}
  \caption{\label{tbl:CNbetan}
    Coefficients $C_{\beta{}N{}n}$ in Eq.~\eqref{eq:xi-integration-formula},
    for $d=3$,
    expanded in $\psi=p_F\ell\gg1$ up to $\mathcal{O}(\psi^{1})$.
  }
  \renewcommand{\arraystretch}{1.5}
  % C_table([0, 1, 2], 4, func=C_tilde, exact=False, order=-1, ndim=3)
\begin{tabular}{c||c|c|c|c|c}\hline\hline$\beta$, $N$ & $n=0$ & $n=1$ & $n=2$ & $n=3$ & $n=4$\\\hline \hline $0$, $1$ & $\infty$ & $1$ & - & - & - \\ $0$, $2$ & $\frac{i\psi}{2}$ & $0$ & $-\frac{i\psi}{2}$ & - & - \\ $0$, $3$ & $-\frac{3 i\psi}{16}$ & $\frac{\psi^{2}}{8}$ & $\frac{i\psi}{16}$ & $-\frac{3 \psi^{2}}{8}$ & - \\ $0$, $4$ & $\frac{i\psi^{3}}{16} -\frac{5 i\psi}{128}$ & $-\frac{\psi^{2}}{32}$ & $-\frac{i\psi^{3}}{16} +\frac{i\psi}{128}$ & $\frac{\psi^{2}}{32}$ & $\frac{5 i\psi^{3}}{16} -\frac{i\psi}{128}$ \\  \hline $1$, $1$ & $\infty$ & $\infty$ & - & - & - \\ $1$, $2$ & $\infty$ & $0$ & $-\frac{i\psi}{2}$ & - & - \\ $1$, $3$ & $-\frac{9 i\psi}{16}$ & $\frac{\psi^{2}}{8}$ & $\frac{3 i\psi}{16}$ & $-\frac{3 \psi^{2}}{8}$ & - \\ $1$, $4$ & $\frac{i\psi^{3}}{16} +\frac{15 i\psi}{128}$ & $-\frac{3 \psi^{2}}{32}$ & $-\frac{i\psi^{3}}{16} -\frac{3 i\psi}{128}$ & $\frac{3 \psi^{2}}{32}$ & $\frac{5 i\psi^{3}}{16} +\frac{3 i\psi}{128}$ \\  \hline $2$, $1$ & $\infty$ & $\infty$ & - & - & - \\ $2$, $2$ & $\infty$ & $\infty$ & $-\frac{i\psi}{2}$ & - & - \\ $2$, $3$ & $\infty$ & $\frac{\psi^{2}}{8}$ & $\frac{5 i\psi}{16}$ & $-\frac{3 \psi^{2}}{8}$ & - \\ $2$, $4$ & $\frac{i\psi^{3}}{16} +\frac{75 i\psi}{128}$ & $-\frac{5 \psi^{2}}{32}$ & $-\frac{i\psi^{3}}{16} -\frac{15 i\psi}{128}$ & $\frac{5 \psi^{2}}{32}$ & $\frac{5 i\psi^{3}}{16} +\frac{15 i\psi}{128}$\end{tabular}
\end{table}

The integral over $z$ can be evaluated by contour integration, with
the help of an analytic function $q(z)$ satisfying a Riemann--Hilbert
problem $q(z+i0^+)-q(z-i0^+)=\theta(1+z)(1+z)^{d/2-1}$ on a branch
cut along the real axis.  The result is given by
\begin{align}
  C_{\beta{}N{}n}
  &=
  -
  2
  i^{n}(-1)^N
  \psi^{-n}
  \sum_{\pm}\Res_{z=\pm i/\psi}
  q(z)
  \frac{
    (1 + z)^{\beta} z^{N-n}
  }{
    (z^2 + \psi^{-2})^N
  }
  \,,
\end{align}
where we can take
\begin{align}
  q(z)
  =
  \begin{cases}
    \frac{i}{2\pi}\ln(-1-z)(1 + z)^{d/2-1}
    \,,
    &
    \text{$d$ even}
    \,,
    \\
    \frac{i}{2}\sqrt{-1-z}(1 + z)^{(d-1)/2-1}
    \,,
    &
    \text{$d$ odd}
    \,.
  \end{cases}
\end{align}
Calculating the residue and expanding the result in series in $\psi$
for $\psi\gg1$ is straightforward with computer algebra.  The values
for $C_{N\beta{}n}$ are shown in Table~\ref{tbl:CNbetan} for $d=3$,
expanded in series of $\psi\gg1$ truncated to order $\psi^{1}$.  To
this order, one can show that the results are the same as from a pole
approximation neglecting the band bottom.

The momentum integral converges for $\beta + d/2 < N + n$, and for
other values the constants are diverging.  Inspection of the gradient
expansion indicates that momentum integrals appearing in order $k>d$
of the expansion are all convergent. For $d=2$ and $3$, a non-convergent
integral appears in the second-order gradient expansion, but can
be removed by requiring that the expansion is gauge invariant. Namely,
for constant scalar $A_i$,
\begin{align}
  0
  &=
  \Tr\ln \mathcal{G}_{\vec{p}+\vec{A}}^{-1}
  -
  \Tr\ln \mathcal{G}_{\vec{p}}^{-1}
  \simeq
  -
  \frac{1}{2m}
  \Tr[\partial_{p_j}(p_i\mathcal{G})] A_i A_j
  \,.
\end{align}
This equation implies a sum rule $C_{120}+\frac{d}{2}C_{010}=-C_{122}$,
using which eliminates all divergent constants appearing.
This operation corresponds to subtraction of the above total derivative.

The above also allows evaluating the $\Omega=\check{A}_\mu=0$ uniform saddle-point
equation of Eq.~\eqref{eq:S-action} exactly. Making an Ansatz $Q=b\Lambda$ with some scalar $b$,
we have
\begin{align}
  \label{eq:b-uniform}
  b \Lambda
  &=
  \frac{i}{\pi\nu_F} \sideset{}{'} \sum_{\vec{p}}
  G(\vec{p})
  =
  \sum_{\vec{p}}
  \frac{
    \frac{1}{2\pi\tau\nu_F}
  }{
    [\mu - \frac{p^2}{2m}]^2 + \frac{b^2}{4\tau^2}
  }
  b \Lambda
  \,,
\end{align}
where the primed sum is defined with the trace part subtracted,
consistent with $\tr \Lambda=0$. Hence,
\begin{align}
  b
  =
  C_{011}\rvert_{\psi\mapsto{}\psi/b}
  =
  \Re\sqrt{1 + \frac{i b}{\psi}}
  \,,
\end{align}
From this, it follows
$b=1+\mathcal{O}(\psi^{-2})$, so with the accuracy we work with, we
can take $b=1$.

\subsection{Noncommutative reduction}

The final step is rewriting the expansion, which consists of terms
with monomials of symbols $\{\Lambda,a_j,a_0\}$, in terms of
gauge-invariant derivatives and the field strength.  We do this using
a similar Gaussian elimination approach as used in noncommutative
Gr\"obner basis constructions. \cite{mora1994}
We outline the approach briefly below.

We define additional symbols $\bar{Q}_i$, $\bar{Q}_{ij}$, $\bar{F}_{ij}$,
and consider the relations
\begin{align}
  \label{eq:Qidef}
  \bar{Q}_i
  &=
  [-ia_i, \Lambda]
  \,,
  \\
  \label{eq:Qijdef}
  \bar{Q}_{ij}
  &=
  [-ia_{ij}, \Lambda] + [-ia_j, [-ia_i, \Lambda]]
  \,,
  \\
  \label{eq:Fijdef}
  \bar{F}_{ij}
  &
  =
  a_{ji} - a_{ij} - i[a_i,a_j]
  \,.
\end{align}
The relations to the actual $Q$ and field strength are then
$\bar{Q}_i=  T^{-1} \hat{\nabla}_iQ T$,
$\bar{Q}_{ij}=T^{-1} \hat{\nabla}_j\hat{\nabla}_iQ T$
and $\bar{F}_{ij}=T^{-1} F_{ij} T$.

Consider now the problem of rewriting an expression $S=\sum_j c_j m_j$
expressed in terms of $a,\Lambda$
solely in terms of $\bar{Q}$, $\bar{F}$, and $\Lambda$.  It is understood the
expression is under $\tr$, and monomials can be cyclically permuted to
their minimal form.  Note that the factors of $T$, $T^{-1}$ cancel
under $\tr$ in expressions containing only the symbols $\bar{Q}$, $\bar{F}$, with
replacement $\Lambda\leftrightarrow{}Q$.

The above relations~(\ref{eq:Qidef}-\ref{eq:Fijdef}) can be expressed as $g_j=0$, $j=1,2,3$,
where
$g_j = \sum_k c_{jk}' m_{jk}'$ are expressions containing symbols $a$,
$\bar{Q}$, $\bar{F}$, $\Lambda$.  Form then the set $I$ of ideal
generators $g_{kj}' = n_k g_j$, where $n_k$ are any monomials such
that the maximum order of terms in $g'$ is $\le{}M$ where $M$ is a
constant; here we can take $M=6$.  We permute all monomials in
$g_{kj}'$ cyclically to their minimal form. The set $I$ has a finite
number of elements. Obviously, each expression $g'$ in $I$ satisfies
$\tr g'=0$ if the definitions of $\bar{Q},\bar{F}$ symbols hold.
Gr\"obner basis algorithms use a more optimal construction of the set
$I$, although they usually don't consider trace permutations.

Define now a monomial ordering so that all monomials $m$ containing
$a$-symbols satisfy $m\succ{}m'$ for all $m'$ not containing any
$a$-symbols. The monomials, ordered from largest to smallest, can be
considered as the basis of a vector space: one can then perform
Gaussian elimination on the ideal generator set $I=\{g'\}$ and the
expression $S$, to find the scalar coefficients $d_j$ which eliminate
the largest terms (in the basis sense) in $S'= S-\sum_{j}d_jg_j'$.
Because $a$-symbols were considered largest, Gaussian elimination
removes them first, and the resulting expression $S'$ will not contain
them if $S$ is gauge-invariant.  The results of the transformation are
straightforward to verify.

The same approach can be used to rewrite results in terms of the
symmetrized gauge-invariant derivatives, by introducing symmetrized
derivative symbols ordered smaller than others, and for other forms
of symbolic simplification of the noncommutative expressions under trace.

\subsection{Results}
\label{app:results}

We list below the typeset output of the program in Supplementary
Material \cite{supplement} that performs the computations outlined
above.  The results are the leading contributions to the local
gradient expanded action, written in form
$\delta S=-\frac{\pi\nu_F}{2}\int\dd{^3r}\tr \bar{S}$
and in units with $\ell=\tau=1$.

The spatial gradient part,
up to order 4 in gradients and to $\psi^{-1}$ in $\psi=p_F\ell$,
\begin{align}
  \bar{S}_{\rm grad}
  &=
  \bar{S}_2
  +
  \bar{S}_{4,0}
  +
  \frac{1}{\psi}
  \bar{S}_{4,1}
  \,,
  \\
\bar{S}_{2} &=  -\frac{1}{12} i{Q}_{i} {Q}_{i} \,,\\
\bar{S}_{4,0} &=  \frac{1}{20} i{Q}_{(ii} {Q}_{jj)}  -\frac{1}{16} i{Q}_{(i} {Q}_{i} {Q}_{j} {Q}_{j)}  \,,\\
\bar{S}_{4,1} &=  -\frac{1}{12}{Q}_{i} {Q}_{j} {Q}_{ij} + \frac{1}{12} i{F}_{ij} {Q} {Q}_{i} {Q}_{j} \,,
\end{align}
Here, $(\ldots)$ in indices means tensor symmetrization,
$Q=T\Lambda{}T^{-1}$,
and $Q_{i}=\hat{\nabla}_iQ$,  $Q_{ij}=\hat{\nabla}_j\hat{\nabla}_iQ$,
and $F$ is the field strength.

The first term in $\bar{S}_{4,1}$ can be written in various forms under trace,
using Eqs.~\eqref{eq:Qidef}, \eqref{eq:Qijdef}.
For example $\Tr({Q}_{i} {Q}_{j} {Q}_{ij})=-\Tr(QQ_{ij}Q_{ij})=-\frac{1}{2}\Tr(Q_{ii}Q_jQ_j)$.

The $a_0$ part, up to fourth order in ``small'' and to $\psi^{-1}$ in
$\psi$:
\begin{align}
  \label{eq:Somegastart}
  \bar{S}_{\Omega}
  &=
  \sum_{j=0}^4(\bar{S}_{\Omega,j,0} + \frac{1}{\psi}\bar{S}_{\Omega,j,1})
  \,,
  \\
\bar{S}_{\Omega,0,0} &= \longmath{ 0 \,,}\\
\bar{S}_{\Omega,1,0} &= \longmath{ C_{010}{\Omega} + {Q} {\Omega} \,,}\\
\bar{S}_{\Omega,2,0} &= \longmath{ -\frac{1}{4} i{Q}_{0} {Q}_{0} \,,}\\
\bar{S}_{\Omega,3,0} &= \longmath{ -\frac{1}{6}{F}_{0i} {Q}_{i} -\frac{1}{4} i{Q} {Q}_{i} {Q}_{0i} -\frac{1}{4} i{Q} {Q}_{0} {Q}_{00} \,,}\\
\bar{S}_{\Omega,4,0} &= \longmath{ \frac{1}{6} i{F}_{0i} {F}_{0i} + \frac{1}{2} i{Q}_{0i} {Q}_{0i} + \frac{1}{4} i{Q}_{00} {Q}_{00} + {F}_{0i} {Q} {Q}_{0i} + \frac{2}{3}{F}_{0i} {Q}_{0} {Q}_{i} + \frac{1}{3}{F}_{0i} {Q}_{i} {Q}_{0} -\frac{5}{16} i{Q}_{0} {Q}_{0} {Q}_{0} {Q}_{0} -\frac{5}{12} i{Q}_{0} {Q}_{0} {Q}_{i} {Q}_{i} -\frac{5}{24} i{Q}_{0} {Q}_{i} {Q}_{0} {Q}_{i} -\frac{1}{6} i{F}_{0i} {Q} {F}_{0i} {Q} \,,}
\end{align}
and
\begin{align}
\bar{S}_{\Omega,0,1} &= \longmath{ 0 \,,}\\
\bar{S}_{\Omega,1,1} &= \longmath{ 0 \,,}\\
\bar{S}_{\Omega,2,1} &= \longmath{ \frac{1}{2}{Q} {\Omega} {\Omega} \,,}\\
\bar{S}_{\Omega,3,1} &= \longmath{ -\frac{1}{4} i{Q}_{0} {Q}_{0} {\Omega} -\frac{1}{4} i{Q}_{i} {Q}_{i} {\Omega} \,,}\\
\bar{S}_{\Omega,4,1} &= \longmath{ -i{F}_{0i} {Q}_{0i} -\frac{1}{2}{F}_{0i} {Q}_{i} {\Omega} + \frac{1}{8}{Q} {Q}_{0i} {Q}_{0i} -\frac{3}{8}{Q}_{0} {Q}_{i} {Q}_{0i} -\frac{1}{2} i{F}_{0i} {Q} {Q}_{0} {Q}_{i} -\frac{1}{4} i{F}_{0i} {Q} {Q}_{i} {Q}_{0} -\frac{3}{4} i{Q} {Q}_{i} {Q}_{0i} {\Omega} -\frac{1}{4} i{Q} {Q}_{0} {Q}_{00} {\Omega} \,,}
  \label{eq:Somegaend}
\end{align}
where we define $Q_0={}[-i\Omega,Q]$ for the ``time derivative'', and
$F_{k0}=-F_{0k}=T(a_{0k}-i[a_k,a_0])T^{-1}=\partial_k\Omega-i[A_k,
  \Omega]=\hat{\nabla}_k\Omega$ for the ``field strength''.  Above, in
contrast to the main text, we set $\check{A}_0=0$. To recover the
$0$ components of the fields and the field strengths in Eqs.~(\ref{eq:Somegastart}-\ref{eq:Somegaend}), shift
$\Omega\mapsto\Omega+\check{A}_0$ so that also
$F_{k0}=-F_{0k}\mapsto{}\hat{\nabla}_k\Omega+\check{F}_{k0}$.

\end{document}